\documentclass[12pt]{iopart}
\usepackage{graphicx}
\usepackage{dcolumn}
\begin{document}

\title{Mean field ground state of a spin-1 condensate in a magnetic field}

\author{Wenxian Zhang, Su Yi and Li You}

\address{School of Physics, Georgia Institute of
Technology, Atlanta GA 30332-0430, USA}

\begin{abstract}
We revisit the topic of the mean field ground state of a spin-1
atomic condensate inside a uniform magnetic field ($B$) under the
constraints that both the total number of atoms ($N$) and the
magnetization ($\cal M$) are conserved. In the presence of an
internal state (spin component) independent trap, we also
investigate the dependence of the so-called single spatial mode
approximation (SMA) on the magnitude of the magnetic field and
${\cal M}$. Our result indicate that the quadratic Zeeman effect
is an important factor in balancing the mean field energy from
elastic atom-atom collisions that are known to conserve both $N$
and $\cal M$.
\end{abstract}



\maketitle

\section{Introduction}
Atomic Bose-Einstein condensates (BEC) have provided a successful
testing ground for theoretical studies of quantum many-body
systems \cite{theory}. In earlier BEC experiments, atoms were
spatially confined with magnetic traps, which essentially freeze
the atomic internal degrees of freedom \cite{scalor}. Most studies
were thus focused on scalar models, i.e. single component quantum
degenerate gases \cite{Dalvo}. More recently, the emergence of
spin-1 condensates \cite{kurn, mike, KetterleNature} (of atoms
with hyperfine quantum number $F=1$) has created opportunities for
understanding degenerate gases with internal degrees of freedom
\cite{Ho98, Law, Pu, Machida, Pu1999, SMA}.

In this paper, we investigate the mean field ground state
structures of a spin-1 atomic condensate in the presence of an external
magnetic field ($B$). We focus on several aspects of the ground
state properties strongly affected by the requirement that
elastic atom-atom collisions conserve both the total number of
atoms ($N$) and the magnetization (${\cal M}$). Several earlier
studies have focused on the global ground state structures when
the conservation of ${\cal M}$ was ignored, or in the limiting
case of a vanishingly small magnetic field ($B=0$)
\cite{KetterleNature, Ho98, Law, Pu, Machida, Pu1999, SMA}.
As we show in this study, in the presence of a
nonzero magnetic field, the conservation
of ${\cal M}$ leads to ground state population distributions
significantly different from that of the global ground state.

Our system is described by the Hamiltonian (repeated
indices are summed) \cite{Ho98}
\begin{eqnarray}
\fl {\cal H} =\int d\vec r\, \psi_i^\dag({\cal L}_{ij}+H_{ZM})
\psi_j +{\frac{c_0}{2}}\int d\vec r
\psi_i^\dag\psi_j^\dag\psi_j\psi_i +{\frac{c_2}{2}}\int d\vec r
\psi_k^\dag\psi_i^\dag\left(F_\eta
\right)_{ij}\left(F_\eta\right)_{kl}\psi_j \psi_l, \label{h}
\end{eqnarray}
where $\psi_j(\vec r)$ is the field operator that annihilates an
atom in the $j$-th ($j=+,0,-$) internal state at location $\vec
r$, ${\cal L}_{ij}\equiv\left[-{\hbar^2\nabla^2}/{2M} +V_{{\rm
ext}}(\vec r) \right]\delta_{ij}$ with $M$ the mass of each atom
and $V_{{\rm ext}}(\vec r)$ an internal state independent trap
potential. Terms with coefficients $c_0$ and $c_2$ of Eq.
(\ref{h}) describe elastic collisions of the spin-1 atom
($|F=1,M_F=+,0,-\rangle$), expressed in terms of the scattering
length $a_0$ ($a_2$) for two spin-1 atoms in the combined
symmetric channel of total spin $0$ ($2$),
$c_0=4\pi\hbar^2(a_0+2a_2)/3M$ and $c_2=4\pi\hbar^2(a_2-a_0)/3M$.
$F_{\eta=x,y,z}$ are spin-1 matrices with
\begin{eqnarray}
F_x={1\over\sqrt{2}}\left(\begin{array}{ccc}0&1&0\\1&0&1\\0&1&0\end{array}\right),
\hskip 6pt F_y={i\over\sqrt{2}}\left(\begin{array}{ccc}0&-1&0\\1&0&-1\\0&1&0\end{array}\right),
\hskip 6pt F_z=\left(\begin{array}{ccc}1&0&0\\0&0&0\\0&0&-1\end{array}\right).
\nonumber
\end{eqnarray}

The external magnetic field $B$ is taken to be along the
quantization axis ($\hat z$), it induces a Zeeman
shift on each atom given by
\begin{eqnarray}
H_{ZM}(B)=\left(%
\begin{array}{ccc}
E_+ & 0 & 0 \\
0 & E_0 & 0 \\
0 & 0 & E_-%
\end{array}%
\right).\nonumber
\end{eqnarray}
According to the Breit-Rabi formula \cite{para1}, the individual level shift
can be expressed as
\begin{eqnarray}
E_+&=&-{E_{\rm HFS}\over 8}-g_I\mu_IB
-{1\over 2}E_{\rm HFS}\sqrt{1+\alpha+\alpha^2},\nonumber \\
E_0&=&-{E_{\rm HFS}\over 8}
-{1\over 2}E_{\rm HFS}\sqrt{1+\alpha^2},\nonumber \\
E_-&=&-{E_{\rm HFS}\over 8}+g_I\mu_IB -{1\over 2}E_{\rm
HFS}\sqrt{1-\alpha+\alpha^2}, \label{br}
\end{eqnarray}
where $E_{\rm HFS}$ is the hyperfine splitting \cite{para1}, and $g_I$
is the Lande $g$-factor for the atomic nuclei with nuclear spin
${\vec I}$. $\mu_I$ is the nuclear magneton and
$\alpha=(g_I\mu_IB+g_J\mu_BB)/E_{\rm HFS}$ with $g_J$ the Lande
$g$-factor for the valence electron with total angular momentum ${\vec
J}$. $\mu_B$ is the Bohr magneton.

\section{Mean field approximation}
\begin{figure}
\begin{center}
\includegraphics[width=4in]{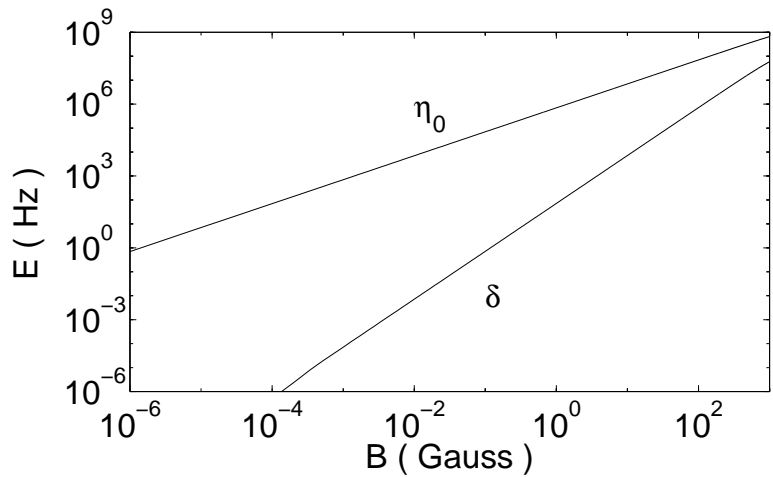}
\end{center}
\caption{Approximate linear and quadratic Zeeman effects as characterized by
parameters $\eta_0$
and $\delta$ versus magnetic field $B$ for a $^{87}$Rb atom.}
\label{eta-delta}
\end{figure}

At near zero temperatures and when the total number of condensed
atoms is large, the ground state is essentially determined by the
mean field term $\Phi_i=\langle\psi_i\rangle$. Neglecting all quantum
fluctuations we arrive at the mean field energy functional from
Eq. (\ref{h}) \cite{Law,spin1}
\begin{eqnarray}
H[\{\Phi_i\}] =H_S+E_0N + {c_2\over 2}\langle\vec
F\rangle^2-\eta_0\langle F_z\rangle+\delta\langle F_z^2\rangle,
\label{mh}
\end{eqnarray}
where the symmetric part
\begin{eqnarray}
H_S=\int d\vec r \left[\Phi_i^* {\cal L}_{ij} \Phi_j +
{\frac{c_0}{2}}\Phi_i^*\Phi_j^*\Phi_j\Phi_i\right],
\end{eqnarray}
is invariant under the exchange of spin component indices,
thus is independent of the external $B$ field. The Zeeman shift as
given by the Breit-Rabi formula (\ref{br}) can be described by two
positive parameters \cite{spin1}
\begin{eqnarray}
2\eta_0 &=& E_--E_+,\nonumber\\
2\delta &=& E_++E_--2E_0,
\end{eqnarray}
which measure approximately the linear and quadratic Zeeman
effects. The $B$-field dependence of
$\eta_0$ and $\delta$ for a $^{87}$Rb atom are displayed in Fig. \ref{eta-delta}.

The elastic atomic collisions as described
by the $c_0$ and $c_2$ parts of the Hamiltonian (\ref{h})
conserve both $N$ and ${\cal M}$,
which in the mean field approximation are given by
\begin{eqnarray}
N &=& \sum_{j=\pm,0}\int d\vec r\, \langle\psi_j^\dag (\vec r)\psi_j(\vec
r)\rangle
\approx \sum_{j=\pm,0}\int d\vec r\,|\Phi_j(\vec r)|^2,\nonumber\\
{\cal M}&=& \int d\vec r\, [\langle\psi_+^\dag (\vec r)\psi_+ (\vec r)\rangle
-\langle\psi_-^\dag (\vec r)\psi_- (\vec r)\rangle]\nonumber\\
&\approx& \int d\vec r\, [|\Phi_+(\vec r)|^2-|\Phi_-(\vec r)|^2].
\label{2cons}
\end{eqnarray}

Before continuing our discussion of the ground state structures,
we shall first briefly comment on the importance of the above two
constraints. In a typical experiment, the last stage before
condensation consists of atomic evaporations, during which neither
$N$ nor ${\cal M}$ is conserved. For a scalar condensate,
typically the ground state is obtained from a minimization of Eq.
(\ref{mh}) subjected to the constraint of only $N$ conservation.
This gives rise to the Gross-Pitaevskii equation (GPE) and the
associated condensate chemical potential, which mathematically is
simply the Lagrange multiplier of the constrained minimization.
A spin-1 condensate requires the introduction of two Lagrange
multipliers during the minimization subjected to both the $N$ and
${\cal M}$ conservation constraints, as was first performed in
\cite{spin1}.

When atomic interactions are ferromagnetic
($c_2<0$ as for $^{87}$Rb atoms) and when the
external $B$-field is negligible, we have shown previously that
the ground state structure is simply a state where all individual atomic
spins are aligned in the same direction \cite{SMA}.
In this case, the conservation of
${\cal M}$ can be simply satisfied by tilting the quantization
axis away from the direction of the condensate spin.
This can always be done if a system described by
(\ref{h}) is rotationally symmetric, and thus contains the
SO(3) symmetry \cite{Ho98}. The presence of a nonzero $B$ field,
on the other hand, breaks the rotational symmetry, [e.g. the
linear Zeeman shift, reduces the SO(3) to SO(2) symmetry], thus
the conservation of ${\cal M}$ has to be included in the
minimization process directly.

The global ground state phase diagram including both linear
and quadratic Zeeman effect was first investigated
by Stenger {\it et al} \cite{KetterleNature}. In this early
study, although the ${\cal M}$ conservation was
included in their formulation, it was not separately discussed,
consequently their results do not easily apply to systems with
fixed values of ${\cal M}$. The ground
state structures as given in Ref. \cite{KetterleNature}
correspond to the actual ground state as realized through a ${\cal M}$
non-conserving evaporation process (e.g. in the presence of a
nonzero $B$-field) that serves as a reservoir for condensate
magnetization. Our study to be presented here, on the other hand,
would explicitly discuss the phase diagram for fixed values of
${\cal M}$, which could physically correspond to experimental
ground states (with/without a $B$-field) due to a ${\cal M}$
conserving evaporation process. Although more limited, as our
results can be traced to linear trajectories of ${\cal M}={\rm
const.}$ in the phase diagram of Ref.
\cite{spin1}, we expect them to be useful, especially in predicting
ground state structures when a ready-made spinor condensate is
subjected to external manipulations that conserve both $N$ and
${\cal M}$.

When atomic interactions are anti-ferromagnetic ($c_2>0$),
the global ground state was first determined to be a total spin
singlet \cite{Law}. More elaborate studies,
including quantum fluctuations, were performed by Ho
and Yip \cite{HoB} and Koashi and Ueda \cite{Ueda}.
Unfortunately, these results \cite{HoB,Ueda} do not correspond
to actual ground states as realized in current experiments,
because of the presence of background magnetic fields.
For instance, the states as found in Ref. \cite{HoB}
are only possible if the magnetic field $B$ is less than
$70 \mu$G at the condensate density as realized in the
MIT experiments \cite{spin1}. The Zeeman shift
(see Fig. \ref{eta-delta}) due to the presence of even
a small magnetic field can overwhelm atomic mean field
interaction and typical atomic thermal energy,
thus if it were not for the
conservation of ${\cal M}$, the ground state
would simply correspond to all atoms condense into the
lowest Zeeman sublevel of $|M_F=1\rangle$.

We now minimize $H$ of Eq. (\ref{mh}) by denoting
$\Phi_j(\vec r)=\sqrt{N_j}\phi_j(\vec r)e^{-i\theta_j}$, with a
real mode function $\phi_j(\vec r)$ ($\int \phi_j^2(\vec r)d\vec
r=1$) and phase $\theta_j$. It is easy to check that the phase
convention of ferromagnetic/anti-ferromagnetic interactions as obtained
previously \cite{Machida} in the absence of a $B$-field still remains true, i.e.
\begin{eqnarray}
\theta_++\theta_--2\theta_0 &=&0, \qquad  c_2<0\ \mbox{ (ferromagnetic)}, \\
\theta_++\theta_--2\theta_0 &=&\pi, \qquad c_2>0\ \mbox{
(anti-ferromagnetic)}.
\end{eqnarray}

\section{Ground state in a homogeneous system}
In a homogeneous system such as a box type trap (of volume ${\cal
V}$), adopting the above phase convention, the resulting
ground state energy functional becomes ($+/-$ for $c_2<0$ and
$c_2>0$ respectively)
\begin{eqnarray}
\fl H[\{N_i\}] =H_S + E_0N + {c_2\over 2\cal V}\left[(N_+-N_-)^2 +
2N_0(\sqrt{N_+}\pm\sqrt{N_-})^2\right] \nonumber\\
\lo -\eta_0(N_+-N_-)+\delta(N_++N_-).
\label{e1}
\end{eqnarray}
Expressing everything in terms of fractional populations and
fractional magnetization $n_i=N_i/N$ and $m={\cal M}/N$, and note
that $n_++n_- = 1-n_0$, $n_+-n_- = m$, Equation (\ref{e1}) becomes
\begin{eqnarray}
\fl {{H}[\{n_i\}] \over N} = {H_S \over N} + E_0
+c\left[(n_+-n_-)^2 + 2n_0(\sqrt{n_+}\pm\sqrt{n_-})^2\right]
\nonumber
\\ \lo -\eta_0(n_+-n_-)+\delta(n_++n_-), \label{sh}
\end{eqnarray}
with an interaction coefficient $c= {c_2N/2\cal V}$,
tunable through a change of condensate density.

We now minimize Eq. (\ref{sh}) under the two constraints
$n_++n_0+n_-=1$ and $n_+-n_-=m$. We restrict our discussion to the
region $-1<m<1$ as the special cases of $m=\pm 1$ are trivial.
Because $H_S$, $E_0$, $c$, $\eta_0$, and $m$ are all constants for
given values of $B$, $N$, and ${\cal V}$, the only part left to be
minimized is
\begin{eqnarray}
{\cal F} &=&2cn_0(\sqrt{n_+}\pm\sqrt{n_-})^2+\delta(n_++n_-).
\label{ef}
\end{eqnarray}

In the special case of $c=0$, Equation (\ref{ef}) reduces to
\begin{eqnarray}
{\cal F} &=&\delta(n_++n_-).
\end{eqnarray}
The ground state is then very simple. When $\delta>0$, which seems
to be always the case for quadratic Zeeman shift, the minimum is
reached by having as large a $n_0$ (thus as small a $n_++n_-$) as
possible, namely
\begin{eqnarray}
n_0 =1-|m|,\qquad n_+ =\left\{\begin{array}{cc}|m|,&m\geq
0\\0,&m<0
\end{array}\right., \qquad n_-=\left\{\begin{array}{cc}0,&m\geq 0\\|m|,&m<0
\end{array}\right..
\end{eqnarray}
When $\delta=0$, we have (in general) three condensate components
with $n_\pm=(1-n_0\pm m)/2$ and $0\le n_0\le 1-|m|$.

\begin{figure}
\begin{center}
\includegraphics[width=6in]{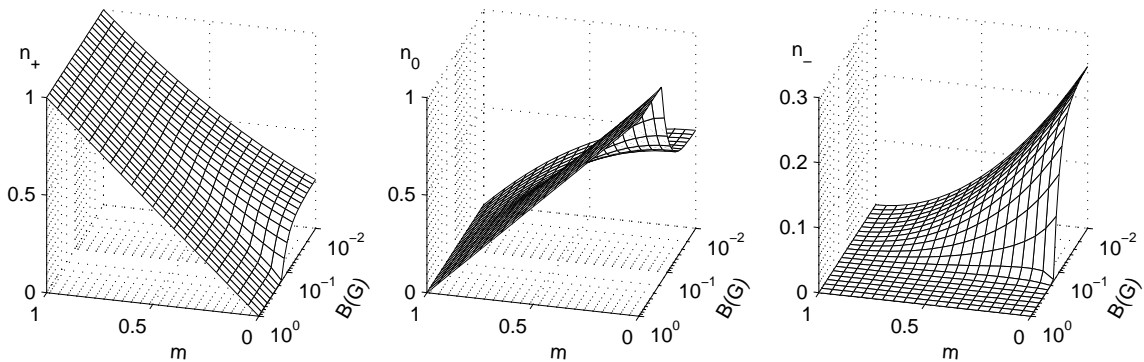}
\end{center}
\caption{The dependence of fractional population for
different spin component on $m$ and
$B$ for a spin-1 $^{87}$Rb homogeneous condensate
with $N/{\cal V}=5\times 10^{14}$cm$^{-3}$.} \label{fig2}
\end{figure}

For ferromagnetic interactions with $c<0$, we define $x=n_++n_-$.
The ground state is then determined by the minimum of
\begin{eqnarray}
{\cal F} &=&g_+(x)+\delta x,
\end{eqnarray}
with $g_+(x)\equiv 2c(1-x)(x+\sqrt{x^2-m^2})$. When $\delta=0$, we
find
\begin{eqnarray}
n_\pm ={1\over 4}(1\pm m)^2,\qquad n_0={1\over 2}(1-m^2),
\label{n0}
\end{eqnarray}
which is the same as obtained in \cite{Pu1999,SMA}. However with a
nonzero $\delta>0$, we find in general
\begin{eqnarray}
n_\pm ={1\over 2}(x_0\pm m),\qquad n_0=1-x_0\ge {1\over 2}(1-m^2),
\label{np0m}
\end{eqnarray}
with $x_0$ being the root of equation $g_+'(x)+\delta=0$, it turns
out that there always exists one and only one solution to the
equation. The equilibrium value for $n_0$ is larger than the
result of Eq. (\ref{n0}) because the quadratic Zeeman effect
causes a lowering of the total energy if two $|M_F=0\rangle$ atoms
are created when an $|M_F=+1\rangle$ atom collides with an
$|M_F=-1\rangle$ atom. Figure \ref{fig2} displays the results of
Eq. (\ref{np0m}) for a typical $^{87}$Rb condensate, for which the
atomic parameters are $E_{\rm HFS}=(2\pi) 6.8347$GHz \cite{para1},
$a_0=101.8a_B$, and $a_2=100.4a_B$ ($a_B$ is the Bohr radius)
\cite{para2}. At weak magnetic fields, typically a condensate
contains all three spin components. With the increasing of
$B$-field, the quadratic Zeeman effect becomes important which
energetically favors the $|0\rangle$ component, so typically only
two components survive: the $|0\rangle$ component and the larger
(initial population) of the $|+\rangle$ or $|-\rangle$ component,
so the ground state becomes (for $m>0$) $n_+ \simeq m$ and $n_0
\simeq 1-m$.

Finally we consider the case of anti-ferromagnetic interactions for $c>0$,
we have then
\begin{eqnarray}
{\cal F} &=&g_-(x)+\delta x, \label{cp}
\end{eqnarray}
with $g_-(x)=2c(1-x)(x-\sqrt{x^2-m^2})$. For $\delta=0$, we again
recover the standard result
\begin{eqnarray}
n_0 =0,\qquad n_\pm = {1\over 2}(1\pm m),
\end{eqnarray}
if $m\ne 0$. When $m=0$, the ground state is under-determined as
many solutions are allowed as along as they
satisfy $n_+=n_-=1-n_0$ with $n_0\in [0,1]$.

In an external $B$-field when $\delta> 0$, we first consider the
special case of $m=0$. It can be easily seen from Eq. (\ref{cp})
that $n_0=1$ is the ground state. For $m\neq 0$, we obtain the
following result: when $\delta>2c[1-\sqrt{1-m^2}]$, the ground
state will have three condensate components with
\begin{eqnarray}
n_\pm=(x_0\pm m)/2, \qquad n_0=1-x_0,
\end{eqnarray}
where $x_0$ is the root of equation $g_-'(x)+\delta=0$; When
$\delta\leq 2c[1-\sqrt{1-m^2}]$, only $|+\rangle$ and $|-\rangle$
components exist, i.e., $n_\pm=(1\pm m)/2$.

\begin{figure}
\begin{center}
\includegraphics[width=6in]{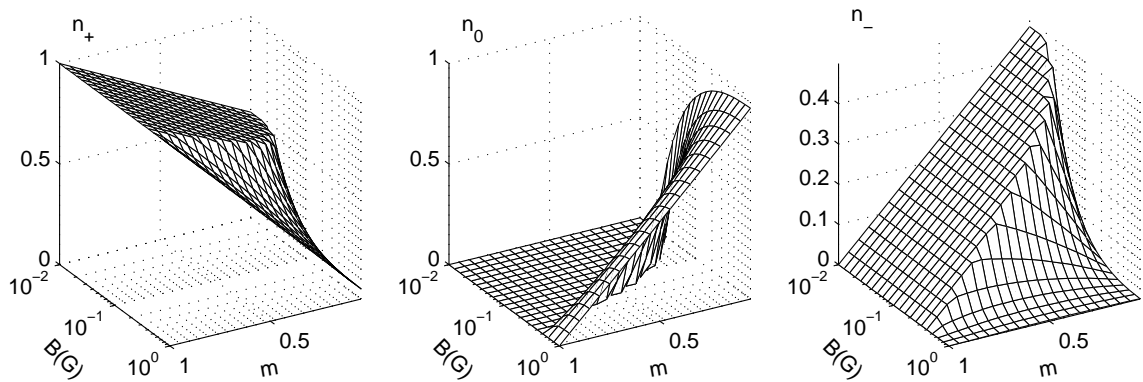}
\end{center}
\caption{The same as in Fig. 2, but now for
a spin-1 $^{23}$Na condensate.} \label{figNah}
\end{figure}

Figure \ref{figNah} is the typical results for a spin-1
$^{23}$Na condensate. The atomic parameters are $E_{\rm
HFS}=(2\pi) 1.7716$GHz \cite{para1}, $a_0=50a_B$ and $a_2=55a_B$
\cite{para3}. At $B=0$ there are only two condensate components,
$|+\rangle$ and $|-\rangle$. For $B>0$ but not very strong, there
are two possibilities: three nonzero condensate components if
$m<m_c$ and two nonzero condensate components if $m\ge m_c$, with
$\delta (B)=2c(1-\sqrt{1-m_c^2}\,)$. When $B$-field gets stronger,
i.e. $\delta(B)\ge 2c$, there are always three condensate
components.

\begin{figure}
\begin{center}
\includegraphics[width=5.5in]{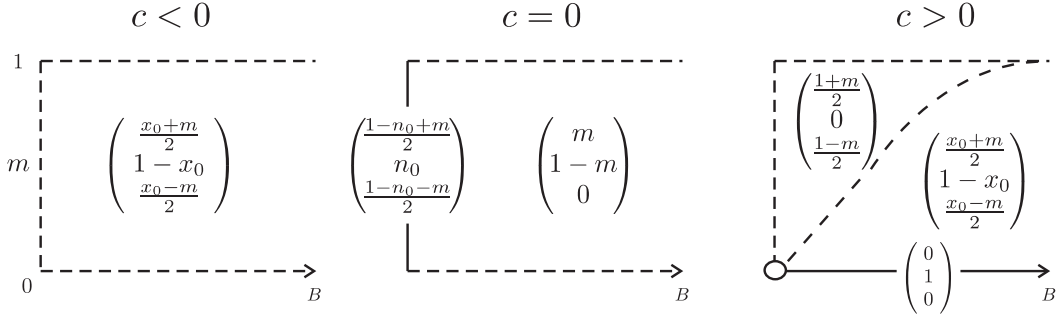}
\end{center}
\caption{The ground state phase diagram for
a homogeneous spin-1 condensate. Dashed curves
and lines denote gradual transitions across the boundaries,
solid lines denote discontinuous jumps. $x_0$ is the solution to equation
$g_\pm'(x)+\delta=0$ and the curves for $c>0$ is determined by
$\delta(B) = 2c[1-\sqrt{1-m^2}]$. The open circle at $B=0, m=0$
for $c>0$ denotes the family of degenerate ground state $\left({1-n_0\over 2},n_0,
{1-n_0\over 2}\right)$.} \label{PhaseDiagram}
\end{figure}

Figure \ref{PhaseDiagram} summarizes the ground state structures
of a homogeneous spin-1 condensate in a $B$-field
for different $c$ and $m$.

\section{Ground state inside a harmonic trap}

In the previous section, we investigated in detail mean
field ground state structures for a spin-1 condensate
in a homogeneous confinement. For the case of a harmonic trap
as in most experiments, there is no reason to believe {\it a priori}
that the above conclusions still hold. In fact, the structures
and phase diagrams
as discussed before is only meaningful if the spatial mode
function $\phi_j(\vec r)$ for different spin components is identical.
Otherwise, it would be impossible to classify the rich variety of
possible solutions. When the spatial mode functions are the same,
the spatial confinement simply introduces an
average over the inhomogeneous density profile of the mode
function.

The aim of this section, is therefore to determine the validity of
the single mode approximation (SMA) in the presence of an external
$B$-field and a harmonic trap. For simplicity, we assume the trap
to be spherically symmetric. We employ numerical methods to
directly find the ground state solutions from the coupled
Gross-Pitaevskii equation
\begin{eqnarray}
i\hbar \frac {\partial}{\partial t}\Phi_+ &=& [{\cal H} + E_+ -
\eta + c_2(n_++n_0-n_-)]\Phi_+ + c_2\Phi_0^2\Phi_-^* \nonumber, \\
i\hbar \frac {\partial}{\partial t}\Phi_0 &=& [{\cal H} + E_0 +
c_2(n_++n_-)]\Phi_0 + 2c_2\Phi_0^*\Phi_+\Phi_- \nonumber, \\
i\hbar \frac {\partial}{\partial t}\Phi_- &=& [{\cal H} + E_- +
\eta + c_2(n_-+n_0-n_+)]\Phi_- + c_2\Phi_0^2\Phi_+^*, \label{3gp}
\end{eqnarray}
subjected to the conservations of both $N$ and
${\cal M}$ [Eqs. (\ref{2cons})]. ${\cal
H}=-\hbar^2\nabla^2/2M+V_{\rm t}(\vec r)+c_0n$, $n_j=|\Phi_j|^2$,
$V_{\rm t}(\vec r)=M\omega^2 r^2/2$, and $n=n_++n_0+n_-$.
$\eta$ is the Lagrange multiplier introduced to numerically enable
the conservation of $\cal M$.

It was shown previously that in the absence of an external
$B$-field, and for ferromagnetic interactions, the SMA is
rigorously valid despite the presence of a harmonic trap
\cite{SMA}. We can also show that in the presence of a nonzero
$B$-field, the linear Zeeman shift does not affect the validity of
the SMA because it can be simply balanced by the external Lagrange
multiplier $\eta$. The quadratic Zeeman effect, on the other hand,
can not be simply balanced, as it favors the production of two
$|0\rangle$ atoms by annihilating one $|+\rangle$ and one
$|-\rangle$ atom during a collision. Such unbalanced elastic
collisions thus break the SO(3) symmetry of the freedom for an
arbitrary quantization axis. Therefore, we do not in general
expect the SMA to remain valid inside a nonzero $B$-field.

Numerically, we find the ground state solutions of Eq. (\ref{3gp}) by
propagating the equations in imaginary time. We typically start
with an initial wave
function as that of a complex Gaussian with a constant velocity:
$\exp[-(x^2/2q_x^2+y^2/2q_y^2+z^2/2q_z^2)-i\vec k\cdot \vec r]$.
$q_x$, $q_y$, $q_z$, and $\vec k$ are adjustable parameters which
are checked to ensure that their choices do not affect the final
converged ground state \cite{SMA}.


For $c_2=0$ or $c=0$, it is easy to check that SMA is always
valid since the energy functional is symmetric with respect to
spin component index. The fractional populations for each
component is therefore the same as for a
homogeneous system, i.e. given by $({1-n_0+m \over 2}, n_0,
{1-n_0-m \over 2})$ if $B=0$, and $(m,1-m,0)$ if $B>0$.

\begin{figure}
\begin{center}
\includegraphics[width=5.5in]{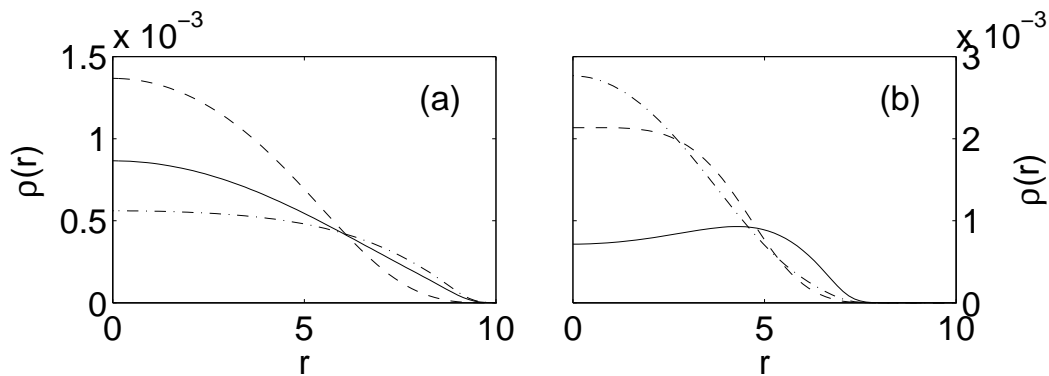}
\caption{Typical densities of spatial mode functions for each
components of a $^{87}$Rb (a) and a $^{23}$Na (b) condensate. The
solid line denotes the $|+\rangle$ component, the dashed line the
$|-\rangle$ component, and the dash-dotted line the $|0\rangle$
component. The parameters are, $N=10^6, \omega = 2\pi\times
100$ (Hz), $B=1.0$ (Gauss), and $m=0.5$.} \label{t0sma}
\end{center}
\end{figure}
\begin{figure}
\begin{center}
\includegraphics[width=5.5in]{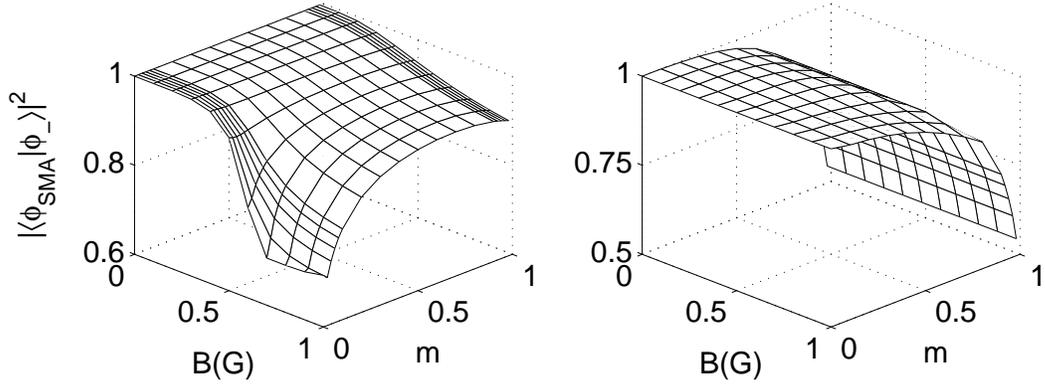}
\caption{The overlap between the SMA mode function and the mode
function for $|-\rangle$ component. Left panel is for a $^{87}$Rb
condensate. Right panel is for a $^{23}$Na condensate.
The atomic parameters are the same as in Fig. 5.}
\label{overlap}
\end{center}
\end{figure}

\begin{figure}
\begin{center}
\includegraphics[width=5in]{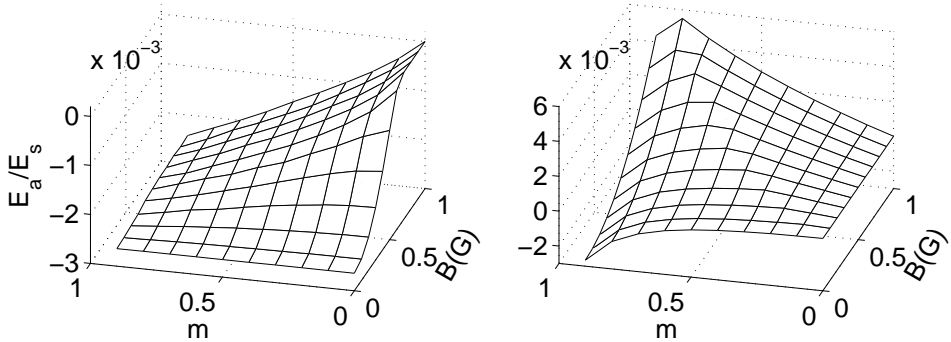}
\caption{The same as in Fig. \ref{overlap},
but now comparing the spin asymmetric energy $E_a=
{c_2}\langle\vec F\rangle^2/2-(\eta_0+\eta)\langle F_z\rangle+\delta\langle F_z^2\rangle $
with the spin symmetric one $H_s$. } \label{ea}
\end{center}
\end{figure}

For $^{87}$Rb and $^{23}$Na condensates, which are believed to be
ferromagnetic $c_2<0$ ($c<0$) and anti-ferromagnetic $c_2>0$
($c>0$) respectively, Figure \ref{t0sma} gives typical density
distributions of spacial mode function, $\rho(\vec r)=|\phi_j(\vec
r)|^2$. Both sub-figures clearly indicate that SMA is no longer
valid. To get an overall idea of the validity of SMA we plot in
Figure \ref{overlap} the overlap integrals of our mode functions
with respect to the SMA mode function $\phi_{\rm SMA}(\vec r)$ as
determined from a scalar GP equation with a nonlinear coefficient
$\propto c_0$ (due to the symmetric $H_S$ only) \cite{SMA}. For a
$^{87}$Rb condensate, we see the overlap is close to unity when
$B$ is small, therefore, SMA remains approximately applicable. But
it becomes increasingly bad with the increase of $B$. We thus
conclude that the SMA remains reasonable in a weak magnetic field
while it is clearly invalid in a strong $B$-field. In fact,
our numerical results confirm that the
stronger the $B$-field, the worse the SMA gets. For typical system
parameters, the dividing line occurs at a $B$-field of a fraction
of a Gauss when the system magnetization ${\cal M}$ is not too
small or too large. For a condensate with anti-ferromagnetic
interactions, it was found earlier that SMA is violated in the
limit of both large $N$ and ${\cal M}$ even without an external
$B$-field, while the case of ${\cal M}=0$ presents an exception
where SMA remains strictly valid for $B=0$ \cite{SMA}. Figure
\ref{overlap} shows the overlap integral for a $^{23}$Na
condensate, indeed we see SMA is invalid except at ${\cal M}=0$
where all atoms are in the $|0\rangle$ component.
Remarkably, despite the seemingly large deviations from
the SMA (as in Fig. \ref{overlap}), the spin
asymmetric energy term remains very small in comparison to
the spin symmetric term as evidenced in Fig. \ref{ea}.

\begin{figure}
\begin{center}
\includegraphics[width=5.5in]{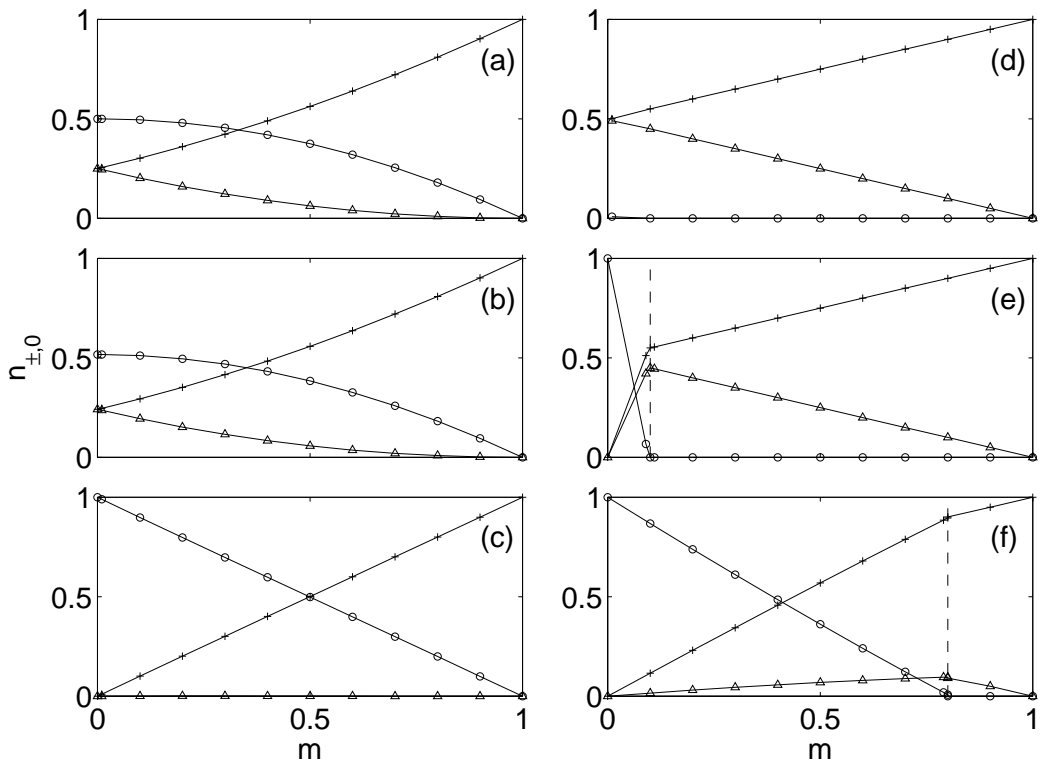}
\caption{Fractional population for each spin component of a $^{87}$Rb
(left column) and a $^{23}$Na (right column) condensate.
The values of $B$-field
from top row to bottom are $B=0,0.1, 1.0$ (Gauss).
The atomic parameters are the same as in Fig. 5.
The solid lines
with plus signs denote the $|+\rangle$ component, the lines with
triangles are for the $|-\rangle$ component and the lines with open
circles for the $|0\rangle$ component. The vertical dashed lines in
(e) and (f) indicate the critical value $m_c$, the boundary between
the two distinct regions discussed in the text. In (d), $m_c=0$.} \label{t0n}
\end{center}
\end{figure}

Figure \ref{t0n} shows the dependence of fractional populations on
the fractional magnetization for a $^{87}$Rb (left column) and
a $^{23}$Na condensate
(right column) at different $B$-fields. For $^{87}$Rb atoms, these
curves resemble the same dependence as for a homogeneous system
where SMA is strictly valid. Nevertheless, we find the
densities of mode functions can become quite different, i.e.
SMA is not valid in general. For $^{23}$Na atoms, the
fractional component populations at different $B$-fields
again follow the results as obtained previously for the homogeneous
case. When $B=0$ [as in Fig. \ref{t0n}(d)], the ground state
distribution clearly obeys the same earlier (homogeneous)
result $({1+m\over 2},0,{1-m \over 2})$, including the
special case when ${\cal M}=0$
where it becomes $({1-n_0 \over 2},n_0,{1-n_0 \over 2})$ with $n_0
\in [0,1]$. For $B \neq 0$ [as in
Fig. \ref{t0n}(e) and (f)], our numerical solutions reveal again
two distinct regions; one for $m < m_c$ where all three components
coexist, and another one for $m
> m_c$ where only two components ($|+\rangle$ and $|-\rangle$)
coexist. We find that $m_c$ increases with the $B$-field, and is
of course limited to $m_c<1$. We conclude that despite the fact a
harmonic trap induces spatially inhomogeneous distribution to
condensate density, thus breaks the SMA in general, the overall ground
state properties as measured by the fractional component
distributions follow closely the results as obtained previously
for the homogeneous case. Physically, we believe the above results
can be understood as fractional populations relate to integrals of
wave functions over all spaces, during which differences between
wave functions can be averaged out. When only
the $|+\rangle$ and $|-\rangle$ components coexist,
in fact, the two constraints on $N$ and
${\cal M}$ always give the fractional population
$n_\pm={(1\pm{m})/2}$ if $N_0=0$.

\begin{figure}
\begin{center}
\includegraphics[width=4in]{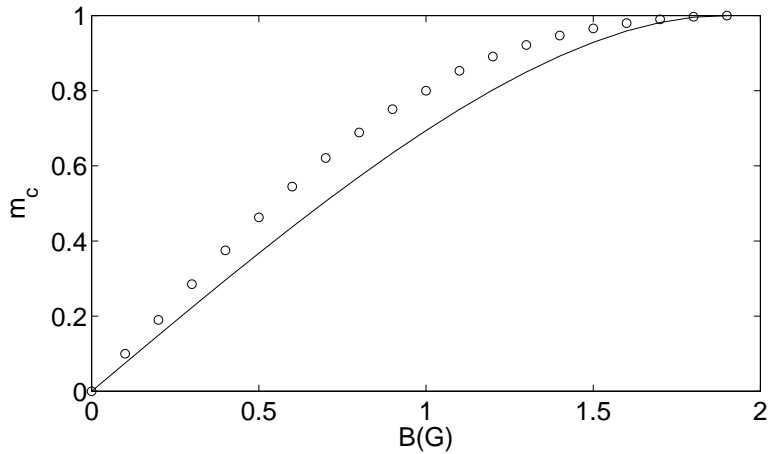}
\caption{The $B$ field dependence of the critical fractional
magnetization $m_c$ as computed numerically for
a $^{23}$Na condensate in a harmonic trap.
The smooth curve corresponds to the result of $\delta=2c[1-\sqrt{1-m_c^2}\,]$
(as from the homogeneous case) with an appropriately adjusted
coefficient $c$ (or density). The atomic parameters are the same as in Fig. 5.}
\label{f8}
\end{center}
\end{figure}

\section{Conclusion}
We have revisited the question of the mean field ground state
structures of spin-1 condensate in the presence of a uniform
magnetic field. For a homogeneous system, when $c=0$, there exists
in general only two nonzero components $|+\rangle$ and
$|0\rangle$, except when $B=0$ where the ground state solution
becomes indefinite; for ferromagnetic interactions when $c<0$, the
ground state in general has three nonzero components; when $c>0$
as for anti-ferromagnetic interactions, except for $m=0$, there
are two regions: one for $\delta>2c[1-\sqrt{1-m_c^2}\,]$ where
three nonzero components coexist and one for $\delta\leq
2c[1-\sqrt{1-m_c^2}\,]$ where only two components coexist. Inside
a harmonic trap, these results remain largely true, although the SMA
becomes generally invalid. We find interestingly (see Fig.
\ref{f8}), the $B$ field (or the $\delta$) dependence of the
critical value $m_c$ that separates the two and three component
condensate regions, remains almost identical as that given by the
analytical formulae $\delta=2c[1-\sqrt{1-m_c^2}\,]$ for a
homogeneous system. In a sense, this also points to
the validity of the use of a mean field description,
as the number of atoms is really large ($10^6$).

\section{Acknowledgement}
We acknowledge interesting discussions with Prof. M. S. Chapman
and Mr. M. -S. Chang. This work is supported by NSF.


\Bibliography{[5]}
\bibitem{theory}See
http://amo.phy.gasou.edu/bec.html/bibliography.html
\bibitem{scalor} Anderson M H, Ensher J R, Matthews M R, Wieman C E
and Cornell E A 1995 {\it Science} {\bf 269} 198; Davis K B, Mewes
M -O, Andrews M R, Druten N J van, Durfee D S, Kurn D M and
Ketterle W 1995 {\it Phys. Rev. Lett.} {\bf 75} 3969; Bradley C C,
Sackett C A, Tollett J J and Hulet R G 1995 {\it Phys. Rev. Lett.}
{\bf 75} 1687; \dash 1997 {\it Phys. Rev. Lett.} {\bf 79} 1170
\bibitem{Dalvo} Dalfovo F, Giorgini S, Pitaevskii L P and
Stringari S 1999 {\it Rev. Mod. Phys.} {\bf 71} 463
\bibitem{kurn} Stamper-Kurn D M, Andrews M R, Chikkatur A P, Inouye
S, Miesner H -J, Stenger J and Ketterle W 1998 {\it Phys. Rev.
Lett.} {\bf 80} 2027
\bibitem{mike} Barrett M D, Sauer J A and Chapman M S 2001 {\it
Phys. Rev. Lett.} {\bf 87} 010404
\bibitem{KetterleNature} Stenger J, Inouye S,
Stamper-Kurn D M, Miesner H -J, Chikkatur A P and Ketterle W 1998
{\it Nature} {\bf 396} 345
\bibitem{Ho98} Ho  T -L 1998 {\it Phys. Rev. Lett.} {\bf 81} 742
\bibitem{Law} Law C K, Pu H, and Bigelow N P 1998 {\it Phys. Rev. Lett.}
{\bf 81} 5257
\bibitem{Pu} Pu H, Law C K, and Bigelow N P 2000 {\it Physica B} {\bf 280} 27
\bibitem{Machida} Ohmi T and Machida K 1998 {\it J. Phys. Soc.
Jpn.} {\bf 67} 1822
\bibitem{Pu1999} Pu H 1999 {\it Phys. Rev. A} {\bf 60} 1463
\bibitem{para1}Vanier J and Audoin C 1988
{\it The quantum physics of atomic frequency standards}
(Philadelphia: A. Hilger)
\bibitem{spin1} Stamper-Kurn D M and Ketterle W 1999
{\it Spinor condensates and light scattering from Bose-Einstein
condensates}, in the Proceedings of Les Houches 1999 Summer
School, Session LXXII, (New York: Springer-Verlag).
\bibitem{SMA} Yi S, M\"{u}stecapl{\i}o\u{g}lu \"{O} E, Sun
C P and You L 2002 {\it Phys. Rev. A} {\bf 66} 011601
\bibitem{para2} Kempen E G M van, Kokkelmans S J J M F, Heinzen D J
and Verhaar B J 2002 {\it Phys. Rev. Lett.} {\bf 88} 093201
\bibitem{para3} Crubellier A, Dulieu O, Masnou-Seeuws F, Elbs M,
Kn\"{o}ckel H and Tiemann E 1999 {\it Eur. Phys. J. D} {\bf 6} 211
\bibitem{HoB} Ho T -L and Yip S K 2000 {\it Phys. Rev. Lett.} {\bf 84}
4031
\bibitem{Ueda} Koashi M and Ueda M 2000 {\it Phys. Rev. Lett.} {\bf 84}
1066
\endbib
\end{document}